# Anomalous Spin and Orbital Hall Phenomena in Antiferromagnetic Systems

J. E. Abrão,[1] E. Santos,[1] J. L. Costa,[1] J. G. S. Santos,[1] J. B. S. Mendes,[2] and A. Azevedo[1]

[1]*Departamento de Física, Universidade Federal de Pernambuco, 50670-901 Recife, Pernambuco, Brasil*
[2]*Departamento de Física, Universidade Federal de Viçosa, 36570-900 Viçosa, Minas Gerais, Brasil*

We investigate anomalous spin and orbital Hall phenomena in antiferromagnetic (AF) materials via orbital pumping experiments. Conducting spin and orbital pumping experiments on YIG/Pt/$Ir_{20}Mn_{80}$ heterostructures, we unexpectedly observe strong spin and orbital anomalous signals in an out-of-plane (OOP) configuration. We report a sevenfold increase in the signal of the anomalous inverse orbital Hall effect (AIOHE) compared to conventional effects. Our study suggests expanding the Orbital Hall angle ($\theta_{OH}$) to a rank 3 tensor, akin to the Spin Hall angle ($\theta_{SH}$), to explain AIOHE. This work pioneers converting spin-orbital currents into charge current, advancing the spin-orbitronics domain in AF materials.

Orbital Hall effect (OHE) provides an intriguing alternative for advancing spintronics, with potential benefits for non-volatile magnetic memories, sensors, microwave oscillators, and nanodevices[1–3]. Recent studies[4–11] have highlighted the significant potential of orbital currents in increasing spin pumping signals driven by ferromagnetic resonance (SP-FMR) and by thermal gradient (spin Seebeck effect (SSE)),[12–14] or in manipulating magnetization through orbital torque.[15–18] However, understanding OHE remains challenging, with research primarily focused on light metals such as Ti, Ru, Cu,[10,12–14] 2D materials,[19–21] and semiconductors.[22] Notably absent are discoveries concerning orbital-to-charge conversion via inverse OHE or inverse orbital Rashba effects in antiferromagnetic (AF) materials, despite their unique properties and increasing interest for spintronic applications. AF materials, characterized by null net magnetization and insensitivity to external magnetic perturbations, exhibit intrinsic high-frequency magnetization dynamics, significant spin-orbit coupling (SOC) and magneto-electric effects. They are recognized as a fertile ground for advanced spintronics research, offering diverse electrical properties and rich opportunities for both experimental investigation and theoretical exploration.[23–30]

In this letter, we investigate the intriguing phenomena of excitation and detection of ordinary and anomalous spin and orbital Hall effects in an AF material. Heterostructures comprising YIG/$Ir_{20}Mn_{80}$(4), YIG/Pt(4), YIG/Pt(2)/$Ir_{20}Mn_{80}$($t$) and YIG/Pt(2)/Ti(4), were utilized, with YIG(400) representing Yttrium Iron Garnet ($Y_3Fe_5O_{12}$) and the AF material consists of $Ir_{20}Mn_8$ (layer thicknesses in nm are indicted in parenthesis). Metallic films were deposited using DC sputtering, and YIG was grown by Liquid Phase Epitaxy (LPE). Measurements were conducted at room temperature using the SP-FMR technique.[31–33] During deposition, the $Ir_{20}Mn_{80}$ films were submitted to a uniform magnetic field (~800 Oe) created by permanent magnets. This procedure aligned the polycrystalline grains inducing an antiferromagnetic texture.[34] To verify the AF phase of the $Ir_{20}Mn_{80}$ film, we performed FMR measurements as a function of the in-plane field in Py(12)/$Ir_{20}Mn_{80}$(15), with Py denoting Permalloy ($Ni_{81}Fe_{19}$). The angular dependence of FMR field exhibited a bell-shaped curve,[35] typical of exchange-biased bilayers, confirming the AF nature of $Ir_{20}Mn_{80}$. Further experimental details can be found in the Supplementary Material (SU) and references[12–14,36].

In the conventional spin-to-charge conversion process using the SP-FMR technique, schematically shown in Fig. 1 (a), an in-plane external field ($\theta = 90°$), pins the YIG magnetization direction. A perpendicular RF field induces uniform magnetization precession under FMR condition, thus inducing the injection of spin accumulation across the interface between the ferromagnet (FM) and the adjacent layer. This accumulation diffuses upward as a spin current $\vec{J}_S$ into the adjacent layer. Through the inverse spin Hall effect (ISHE),[37–41] it generates a perpendicular charge current ($\vec{J}_C$), governed by,

$$\vec{J}_C = (2e/\hbar)\theta_{SH}(\hat{\sigma}_S \times \vec{J}_S), \tag{1}$$

where $\theta_{SH}$ is the spin Hall angle, a constant that measure the efficiency of the spin-to-charge conversion, $\vec{J}_S$ is the spin current direction and $\hat{\sigma}_S$ is the spin polarization direction. The charge current created by the ISHE process is given by $I_{SP-FMR} = V_{SP-FMR}/R$, where $V_{SP-FMR}$ and $R$ are the voltage and electrical resistance directly measured between the electrodes, respectively.

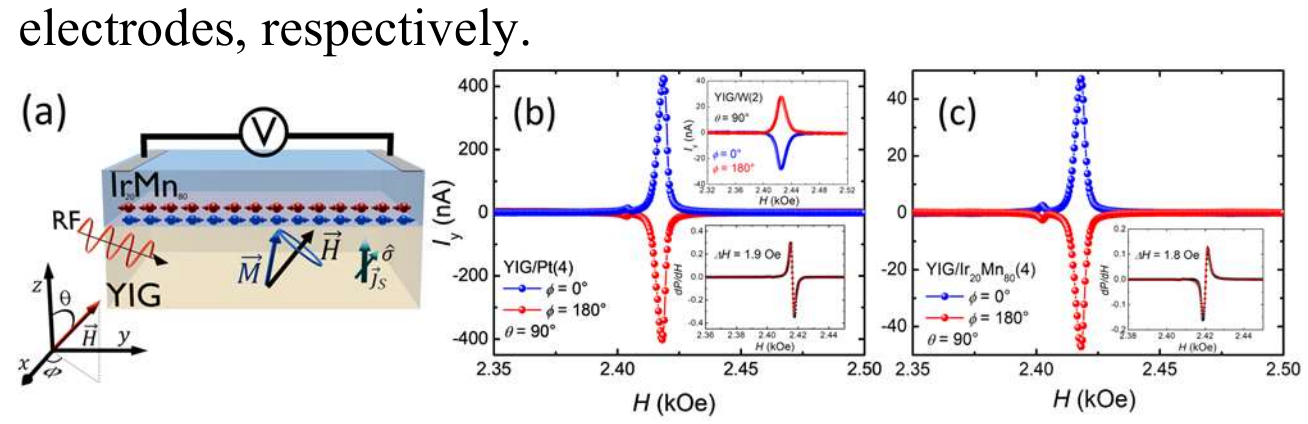

**Figure 1.** (a) Experimental setup for conventional spin pumping measurements, with the external field applied in-plane, ($\theta = 90°$) with $\phi = 0°$ or $\phi = 180°$. (b) SP-FMR signals for YIG/Pt(4). The upper inset shows SP-FMR results for YIG/W, with opposite polarity to YIG/Pt. The lower inset shows the FMR absorption curve for YIG/Pt, with a linewidth of $\Delta H = 1.9$ Oe. (c) SP-FMR signals for YIG/$Ir_{20}Mn_{80}$(4). The inset shows the FMR absorption curve for YIG/$Ir_{20}Mn_{80}$(4), with a linewidth of $\Delta H = 1.8$ Oe. FMR data were performed at an RF power of 6 μW, while SP-FMR measurements used an RF power of 13.4 mW at a frequency of 9.41 GHz.

Fig. 1 (b) shows typical SP-FMR for YIG/Pt(4) in the in-plane configuration. At $\theta = 90°$ and $\phi = 0°$, a positive voltage peak (blue symbols) is detected at the FMR condition. Reversing $\vec{H}$ or rotating the sample to $\phi = 180°$, changes the sign of $\hat{\sigma}_S$, while $\vec{J}_S$ remains unchanged, resulting in a change in the polarity of the measured signal (red symbols), but keeping the magnitude constant. The lower inset shows the derivative of the absorption signal, with an FMR linewidth of 1.8 Oe. The upper inset shows the SP-FMR signal for YIG/W(2), which has opposite polarity to YIG/Pt. Fig.1(c), shows the SP-FMR signal for YIG/$Ir_{20}Mn_{80}$(4). Notably, the ISHE in $Ir_{20}Mn_{80}$ has the same polarity as in Pt, indicating that the SOC in $Ir_{20}Mn_{80}$ is positive, i.e., $\vec{L} \cdot \vec{S} > 0$. However, since the signal magnitude is smaller than for Pt, we conclude that $SOC_{Pt} > SOC_{Ir_{20}Mn_{80}}$.

It is important to mention that the spin current is actually a rank 2 tensor, but for practical purposes, it is useful to decompose this tensor into two distinct physical components: its direction and its polarity. While this separation is mainly motivated for convenience, it is crucial interpreting experimental data. In a typical SP-FMR setup, the spin current ($\vec{J}_S$) always flows out of the FM material, causing spin accumulation that diffuses through the adjacent layer. The spin polarization vector $\hat{\sigma}_S$, however, always aligns with the external magnetic field $\vec{H}$. Notably, the ISHE is independent on the magnetic order of the spin-to-charge converter material.[42,43] The interconversion of spin to charge via spin Hall effects is driven by scattering events within the material's bulk via spin-orbit interactions, whether intrinsic or extrinsic.[38,39]

In recent years, groundbreaking theoretical study[44] has predicted the emergence of anomalous direct and inverse spin Hall effect (ASHE and AISHE, respectively). This significant advance was achieved by extending the conventional spin Hall angle ($\theta_{SH}$) to a rank 3 tensor, taking in account an order parameter in the material of interest. In FM materials, this order parameter can be the magnetization $\vec{M}$, while in antiferromagnetic materials it corresponds to the Néel vector $\vec{n}$. The proposed rank 3 spin Hall angle $\theta_{ijk}^{SH}$ can be defined as:

$$\theta_{ijk}^{SH} = \theta_0 \epsilon_{ijk} + \theta_1 n_i \epsilon_{iln} \epsilon_{jnk} + \theta_2 n_i \epsilon_{ink} \epsilon_{jln}, \quad (2)$$

where $\theta_0$ represents the conventional spin Hall angle used in SHE/ISHE, while $\theta_1$ and $\theta_2$ are the anomalous spin Hall angles. The indexes $i, j, k = 1,2,3$ correspond to the $\hat{x}$, $\hat{y}$ and $\hat{z}$ directions, respectively, with $\varepsilon_{ijk}$ representing the Levi-Civita symbol. Consequently, by expanding the spin Hall angle into a rank 3 tensor, the $\vec{J}_S$ and $\vec{J}_C$ generated via SHE and ISHE gain an additional term which depends on the order parameter:

$$J_k^C = \sum_{ij} \left(\frac{2e}{\hbar}\right) \theta_{ijk}^{SH} j_S^{ij} \text{ and } j_S^{ij} = \sum_k \left(\frac{\hbar}{2e}\right) \theta_{ijk}^{SH} J_k^C, \quad (3)$$

where $J_k^C$ is the charge current applied/detected along a specific $\hat{k}$ direction and $j_S^{ij}$ is the spin current, a rank two tensor where the first index denotes the spin flow direction, and the second index denotes the $\hat{\sigma}_S$ direction.

It is noteworthy that the spin Hall angle, now represented as a rank 3 tensor, enables spin-to-charge conversion even when the spin polarization aligns parallel to the spin flow direction. This scenario is particularly intriguing because any observed signal can be explained by ISHE alone. For example, if we align the vectors $\vec{J}_S$ and $\hat{\sigma}_S$ along the $\hat{z}$ axis, the converted spin current will generate an electrical signal, which follows:

$$J_k^C = (2e/\hbar)(\theta_1 + \theta_2)\delta_{ki\neq 3}\, n_i j_S^{33}. \quad (4)$$

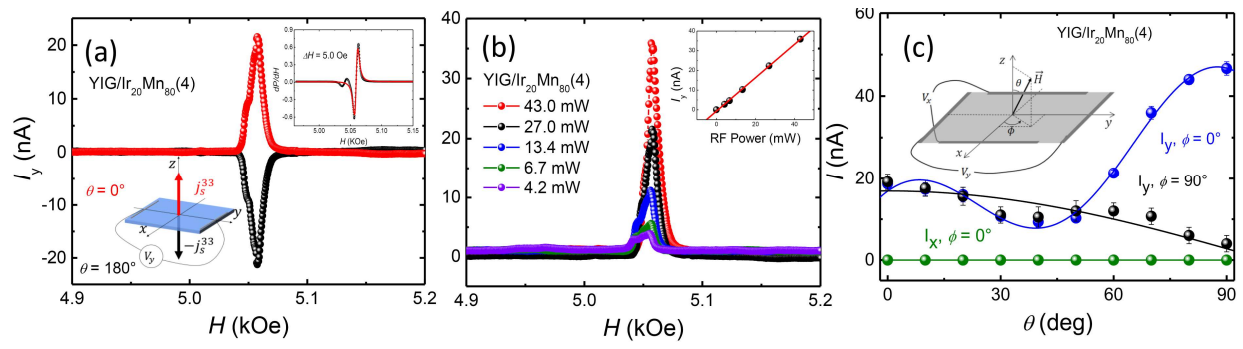

**Figure 2**. (a) OOP SP-FMR signal for YIG/$Ir_{20}Mn_{80}$(4) at $\theta = 0°$ (red) and $\theta = 180°$ (black), with an RF power of 27 mW and frequency of 9.51 GHz. $\theta$ is the polar angle defined in Fig.1(a). The upper inset shows the corresponding FMR signal in the OOP configuration at an RF power 6 μW. The lower inset shows OOP configuration, where $\theta = 0°$ corresponds to $j_S^{33}$, and $\theta = 180°$ corresponds to $-j_S^{33}$. (b) SP-FMR signal at different RF power levels. The inset shows the peak current versus RF power, exhibiting linear behavior. The high quality of our YIG films allows detection of a surface magnetostatic mode below the FMR field, causing broadening of the FMR linewidth in the SP-FMR signals. (c) Polar angle dependence of the SP-FMR peak signal, measured at an RF power of 27 mW and RF frequency of 9.51 GHz. Signals were measured along the Y-direction with the field sweeping in the z-x plane (blue symbols), and z-y plane (black symbols), according to the upper inset. The solid lines were obtained by means of the theory developed in Ref. [44], and detailed in section 1.6 of SM.[36]

Which implies that if the order parameter has components in the x-y plane, a detectable signal can be observed. This result is significant as it introduces the possibility of generating charge current along arbitrary directions, a phenomenon not previously anticipated in conventional ISHE studies.

To explore the AISHE in antiferromagnets, YIG/$Ir_{20}Mn_{80}$(4) samples were fabricated. While the traditional ISHE is investigated by applying an in-plane magnetic field $\vec{H}$, AISHE is investigated by applying the $\vec{H}$ in the OOP configuration, $\theta = 0°$ or $\theta = 180°$. In this setup, the $\vec{J}_S$ direction will be parallel to $\hat{\sigma}_S$, meaning that we are effectively exploring the $j_S^{33}$ component of the spin current tensor, as illustrated in Fig. 2 (a).

Fig. 2(a) shows the SP-FMR signal in the OOP configuration for YIG/$Ir_{20}Mn_{80}$(4) measured along the Y-direction. A well-defined current peak is detected at around 5.05 kOe, corresponding to the excitation of FMR, as shown in the inset. Since the directions of $\hat{\sigma}_S$, and $\vec{J}_S$ are parallel, the measured signal cannot be attributed to the conventional ISHE, described by the Eq. (1). Moreover, due to the insulating nature of YIG, no anomalous Nernst effect or other galvanomagnetic are present. On the other hand, the measured signal fits perfectly with the AISHE as the Néel vector is along the Y-direction. Upon rotating the sample to $\theta = 180°$, the orientation of $\hat{\sigma}_S$ changes, resulting in an inversion of the measured signal. This result differs from previous findings where a FM was used instead of an AF[45]. In the referenced study,[45] it was observed that changing the direction of $\vec{H}$ had no effect on the detected signal. This was attributed to the order parameter (magnetization) that changes exclusively in the sample plane. In the YIG/$Ir_{20}Mn_{80}$ system, the Néel vector serves as the order parameter, which exhibits much stronger rigidity compared to the magnetization of a ferromagnet, thus remaining unaffected by $\vec{H}$ on the order of a few kOe. We also observed that the measured signal responds linearly to the microwave power used to excite the FMR condition as shown in Fig. 2(b). This result indicates that the detected signal depends linearly on the spin current, further supporting the AISHE interpretation. Fig. 2(c) shows the polar angle dependence of the SP-FMR signal for three different configurations used to explore the AISHE. The data along the Y-direction ($V_y$) vary depending on whether the field sweeps in the z-y plane (blue symbols) or the z-x plane (black symbols): (i) for z-y plane ($\phi = 90°$, black symbols), the signal shows no contribution from ISHE; (ii) in the z-x plane ($\phi = 0°$, blue symbols), the signal exhibits a contribution from ISHE and AISHE. The minimum at around 40° results from the completion between ISHE and AISHE, as discussed in the SM.[36] In contrast, measurements along X-direction ($V_x$) show no ISHE or AISHE signal. The solid lines represent theoretical fits based on the model developed X. R. Wang [44], which predicts a charge current $J_c$ that does not follow equation 1. Details of the best fits to the experimental data are discussed in section 1.6 of the SM [36].

Another attractive approach to explore spin-to-charge conversion in AF involves examining orbital effects. In recent years, orbital angular momentum has attracted significant attention due to its ability to impact transport properties, given that non-equilibrium orbital momentum does not suffer quenching.[4] However, experimental studies in antiferromagnets remain scarce,[46–48] with no reports to date on orbital-to-charge conversion via inverse orbital Hall or inverse orbital Rashba effects in this class of materials.

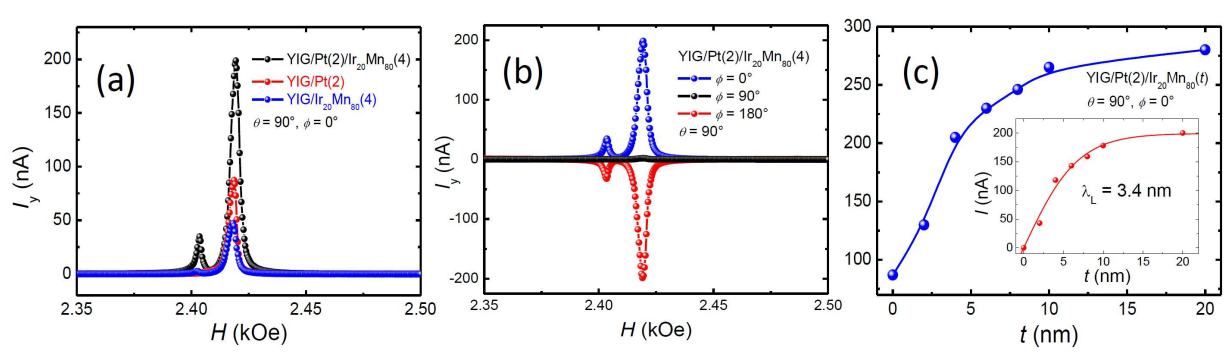

**Figure 3.** (a) SP-FMR signals for YIG/Pt(2)/$Ir_{20}Mn_{80}$(4) (black symbols), YIG/Pt(2) (red symbols), and YIG/ $Ir_{20}Mn_{80}$(4) (blue symbols) in the in-plane configuration ($\theta = 90°$ and $\phi = 0°$). Measurements were performed with an RF power of 13.4 mW, and frequency of 9.41 GHz. The weak SP-FMR signal generated by the surface mode is hardly detected in YIG/$Ir_{20}Mn_{80}$ and YIG/Pt, but shows a significant gain in YIG/Pt(2)/$Ir_{20}Mn_{80}$(4). (b) SP-FMR signal for YIG/Pt(2)/$Ir_{20}Mn_{80}$(4), at $\theta = 90°$, $\phi = 0°$(blue), $\phi = 90°$(black) and $\phi = 180°$ (red). (c) SP-FMR peak signals, for $0nm \leq t_{IrMn} \leq 20nm$. The solid line is to guide the eyes. The inset represents the IOHE for $Ir_{20}Mn_{80}$ films, with the solid red line obtained as discussed in the text.

Based on previous investigation[12–14] we fabricated samples of YIG/Pt(2)/$Ir_{20}Mn_{80}$($t$) with varying thicknesses of the $Ir_{20}Mn_{80}$ layer ranging from $t = 0$ nm to $t = 20$ nm. The YIG/Pt(2) bilayer exhibits two notable characteristics: first, due to the low SOC of YIG, it exclusively injects spin current into Pt. Second, due to the large SOC of Pt, a fraction of the injected spin current undergoes conversion to a charge current via ISHE, while most of the spin current transforms into an entangled spin-orbital current. This entangled spin-orbital current serves as a valuable tool for probing orbital effects in different materials.

Fig. 3(a) shows the SP-FMR signal for YIG/Pt(2), YIG/$Ir_{20}Mn_{80}$(4) and YIG/Pt(2)/ $Ir_{20}Mn_{80}$(4), measured in the in-plane configuration. Direct comparison of the measured signals for the first two samples confirms the larger SOC in Pt compared to $Ir_{20}Mn_{80}$. However, adding a 4nm layer of $Ir_{20}Mn_{80}$ on top of the Pt layer almost doubles the signal compared to ISHE in YIG/Pt. This observed increase cannot be attributed solely to ISHE in $Ir_{20}Mn_{80}$, so orbital Hall effect must be considered. The result in Fig. 3(a) can be explained by analyzing the spin Hall conductivity $\sigma_{SH}$ and the orbital Hall conductivity $\sigma_{OH}$. First principles calculations[49] reveal that Ir has $\sigma_{OH}^{Ir} \sim 4334\ (\hbar/e)(\Omega \cdot \mathrm{cm})^{-1}$ and $\sigma_{SH}^{Ir} \sim 321\ (\hbar/e)(\Omega \cdot \mathrm{cm})^{-1}$, while Mn has $\sigma_{OH}^{Mn} \sim 6087\ (\hbar/e)(\Omega \cdot \mathrm{cm})^{-1}$ and $\sigma_{SH}^{Mn} \sim -37\ (\hbar/e)(\Omega \cdot \mathrm{cm})^{-1}$. Therefore, $Ir_{20}Mn_{80}$ is anticipated to exhibit a strong $\sigma_{OH}$, consequently contributing to a strong SP-FMR signal due to IOHE in $Ir_{20}Mn_{80}$ thin films.

In Fig. 3(b), we present the angular dependence of IOHE in YIG/Pt(2)/$Ir_{20}Mn_{80}$(4). Upon rotating the sample, the signal changes according to an equation analogous to that of ISHE, described mathematically described by:

$$\vec{J}_C = (2e/\hbar)\theta_{OH}(\hat{\sigma}_L \times \vec{J}_L), \tag{5}$$

where $\theta_{OH}$ is the orbital analogous of the $\theta_{SH}$, and measures the orbital-to-charge conversion efficiency while $\hat{\sigma}_L$ represents the orbital polarization. In our approach, the orientation of $\hat{\sigma}_L$ is determined by the spin polarization $\hat{\sigma}_S$ injected into the Pt film via the SP-FMR technique. Since the SOC in Pt is positive, $\vec{L} \cdot \vec{S} > 0$ and $\hat{\sigma}_S || \hat{\sigma}_L$. The thickness dependence in Fig. 3(c) indicates typical diffusion-like behavior, with the signal saturating for thicker films due to the information loss from dissipation mechanisms within the film. The effective charge current from SP-FMR includes contributions from orbital and spin effects for both Pt(2) and $Ir_{20}Mn_{80}$($t$), given by

$$\vec{J}_C^{eff} = (2e/\hbar)[\theta_{SH}^{Pt}(\hat{\sigma}_S \times \vec{J}_S^{Pt}) + \theta_{OH}^{Pt}(\hat{\sigma}_L \times \vec{J}_L^{Pt}) + \theta_{SH}^{IrMn}(\hat{\sigma}_S \times \vec{J}_S^{IrMn}) + \theta_{OH}^{IrMn}(\hat{\sigma}_L \times \vec{J}_L^{IrMn})]. \tag{6}$$

The charge current due to ISHE+IOHE in $Ir_{20}Mn_{80}$, represented in the inset of Fig. 3(c), is given by $I_C^{IrMn} = I_C^{eff} - I_C^{Pt(2)}$, where $I_C^{Pt(2)} = 87$ nA, extracted from Fig. 3(a). The theoretical fit, $I_C^{IrMn} = A \times tanh(t/2\lambda_{LS})$, yields an spin-orbital diffusion length $\lambda_{LS}^{IrMn} = (3.4 \pm 0.5)$ nm, which is greater than the spin diffusion length in Pt, $\lambda_S^{Pt} \sim 1.6$ nm.[12]

To confirm that the enhancement in the signal is due to the orbital-to-charge conversion within the AF layer, we grew YIG/W(2nm) and YIG/W(2nm)/Ir20Mn80 samples. The in-plane SP-FMR signals for these samples are shown in Fig. S11 of the SM. An ISHE signal of -27.9 nA was measured in YIG/W(2) (blue symbols in Fig. S11(a)), consistent with the expected opposite sign compared to YIG/Pt. When a layer of $Ir_{20}Mn_{80}$ was added on top of the W layer, the signal peak increased to -78.3 nA (violet symbols in Fig. S11(a)). This increase is attributed to the IOHE in the $Ir_{20}Mn_{80}$ layer, while the negative value results from the negative spin Hall conductivity of W,[49] turning $\vec{L} \cdot \vec{S} < 0$. This indicates that the orbital current injected from W to $Ir_{20}Mn_{80}$ has the opposite sign compared to the orbital current injected from Pt to $Ir_{20}Mn_{80}$.

Finally, each spin-to-charge conversion mechanism has a corresponding orbital counterpart, originating from different physical mechanisms but producing similar results. This raises the question of whether an Anomalous Inverse Orbital Hall effect (AIOHE) also exist. To explore AIOHE, we conducted experiments using YIG/Pt(2)/$Ir_{20}Mn_{80}$($t$) and YIG/W/ $Ir_{20}Mn_{80}$ samples arranged in the OOP configuration and performed SP-FMR measurements. Fig. 4(a) shows the SP-FMR signal for YIG/$Ir_{20}Mn_{80}$(4) and YIG/Pt(2)/$Ir_{20}Mn_{80}$(4) samples. The peak signal of the YIG/$Ir_{20}Mn_{80}$ sample was around 37.5 nA, while the YIG/Pt(2)/$Ir_{20}Mn_{80}$(4) sample exhibited a significantly higher peak value of 271.6 nA, an increase of more than sevenfold. This surprising increase suggests the existence of an extra spin-orbital to charge conversion mechanism beyond the traditional ISHE or IOHE, given the experimental setup employed. Moreover, the signal cannot be attributed to the AISHE within the Pt layer, as no order parameter exists. When the sample is rotated 180º, the polarity of the signal changes indicating that the measured signal depends on the $\hat{\sigma}_S$ direction. Moreover, it behaves similarly to what was previously observed for the AISHE in YIG/$Ir_{20}Mn_{80}$(4). This suggests that the signal depends on the order parameter of the AF layer, which remains fixed within the applied magnetic field range. This hypothesis is supported by analyzing the SP-FMR signal of YIG/Pt(2)/Ti(4) in the OOP configuration, where no signal is observed, as shown in the inset of Fig. 4(b). Previous experimental results have shown that Titanium is an excellent material to convert orbital current into charge current via IOHE.[13] However, Ti does not exhibit an order parameter, leading to the absence of additional charge current via AIOHE.

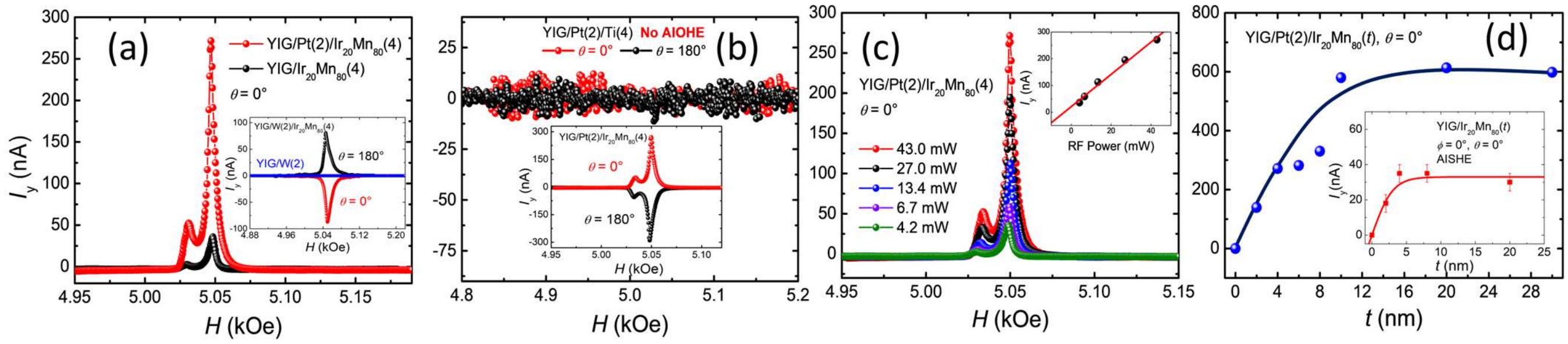

**Figure 4.** (a) SP-FMR signal in the OOP configuration for YIG/$Ir_{20}Mn_{80}$(4) (black symbols) and YIG/Pt(2)/$Ir_{20}Mn_{80}$(4) (red symbols), with an RF power of 43 mW, at 9.51 GHz. The inset shows the SP-FMR signals for YIG/W(2) (blue symbols), and YIG/W(2)/$Ir_{20}Mn_{80}$(4) for $\theta = 0^o$ (red symbols), and $\theta = 180^o$ (black symbols). The increased signal can only be explained as due to the AIOHE. (b) SP-FMR signal for YIG/Pt(2)/Ti(4) with an RF power of 43 mW. As Ti has no order parameter, no AIOHE signal is detected. The inset shows angular dependence of the AIOHE measured in YIG/Pt(2)/$Ir_{20}Mn_{80}$(4). Due to the rigidity of the Néel vector, which acts as the order parameter, the AIOHE signal inverts its polarity at $\theta = 180^o$ compared to $\theta = 0^o$. (c) Dependence of the AIOHE signal on RF power. (d) Dependence of the AIOHE signal on $Ir_{20}Mn_{80}$ layer thickness, with an RF power of 43 mW. The inset shows the dependence of the AISHE signal on $Ir_{20}Mn_{80}$ layer thickness, which is negligible compared with the AIOHE signal.

To further elucidate the behavior of the measured signals, we conducted experiments varying the microwave power. The results, presented in Fig. 4(c), reveal a notable trend: the SP-FMR signal increases as we increase the microwave power. This result indicates a direct correlation between the magnitude of the spin-orbital current injected into the $Ir_{20}Mn_{80}$ material and the observed effect. Plotting the peak signal as function of the microwave power we observed a linear dependence, as shown in the inset of Fig. 4(c). Furthermore, we studied the dependence of AIOHE on $Ir_{20}Mn_{80}$ film thickness. As shown in Fig. 4(d), the signal intensity saturates for thicker films, indicating diffusion-like behavior, due to dissipation mechanism, similar to AISHE measurements. The inset shows the AISHE versus thickness of $Ir_{20}Mn_{80}(t)$ layer measured in YIG/$Ir_{20}Mn_{80}(t)$. The AISHE saturation value (red symbols) is negligible compared to the AIOHE saturation value IOHE (blue symbols).

In summary, our findings present compelling evidence of spin and orbital anomalous Hall signals discovered through SP-FMR experiments in an antiferromagnetic material. This signal, attributed to the Anomalous Inverse Orbital Hall effect, emerged from spin and orbital pumping experiments conducted at room temperature. Unlike conventional ISHE and IOHE, this signal demonstrates unique characteristics dependent on various parameters, including the Néel vector of the AF material, spin and orbital pumping configurations, external magnetic field, and AF layer thickness. Comparing the signals obtained from YIG/Pt(2)/$Ir_{20}Mn_{80}$(4) and YIG/$Ir_{20}Mn_{80}$(4) heterostructures revealed a remarkable seven-fold increase in the AIOHE signal. Just as $\theta_{SH}$ can be expanded to a rank 3 tensor if the converting layer has an order parameter, the $\theta_{OH}$ must also be a rank 3 tensor. By accounting for possible anomalous signals due to the order parameter, the direct and inverse orbital Hall effect will have additional terms to generated/convert orbital currents, analogous to their spin counterpart. Thus, the emergence of the extra signal can be simply explained by the existence of an AIOHE. Although the first-principle explanation for the AISHE and AIOHE is still under theoretical development, we can assert that the breaking of time-reversal symmetry through an order parameter induces a nonzero Berry curvature, which directly contributes to the anomalous signals observed in this work. To date, no other work has explored this pathway to convert spin-orbital currents into charge current, expanding the understanding of spin-orbitronics phenomena.

The authors would like to thank D. S. Maior for helping with the MOKE measurements and with the images. This research is supported by Conselho Nacional de Desenvolvimento Científico e Tecnológico (CNPq), Coordenação de Aperfeiçoamento de Pessoal de Nível Superior (CAPES) (Grant No. 0041/2022), Financiadora de Estudos e Projetos (FINEP), Fundação de Amparo à Ciência e Tecnologia do Estado de Pernambuco (FACEPE), Universidade Federal de Pernambuco, Multiuser Laboratory Facilities of DF-UFPE, Fundação de Amparo à Pesquisa do Estado de Minas Gerais (FAPEMIG) - Rede de Pesquisa em Materiais 2D and Rede de Nanomagnetismo, and INCT of Spintronics and Advanced Magnetic Nanostructures (INCT-SpinNanoMag), CNPq 406836/2022-1.